\documentclass[aps,pra,showpacs,twocolumn,superscriptaddress]{revtex4}

\usepackage{graphicx,color}
\usepackage{amsmath,amsthm,amsfonts,amssymb,bm}
\usepackage{times}
\usepackage{epsf}
\usepackage[colorlinks={true}]{hyperref}

\hypersetup{citecolor={blue}, filecolor={blue}, linkcolor={blue}, urlcolor={blue}}

\newcommand{\commentold}[1]{}

\begin{document}

\title{Shielding quantum discord through continuous dynamical decoupling}

\author{Felipe F. Fanchini}\email{fanchini@fc.unesp.br}
\affiliation{Faculdade de Ci\^encias, UNESP - Univ Estadual Paulista, Bauru, SP, 17033-360, Brazil}
\author{Emanuel F. de Lima}
\affiliation{Instituto de Geoci\^encias e Ci\^encias Exatas, UNESP- Univ Estadual Paulista, Rio Claro, SP, 13506-900, Brazil}
\author{Leonardo K. Castelano}
\affiliation{Departamento de F\'{\i}sica, Universidade Federal de S\~ao Carlos, S\~ao Carlos, SP, 13565-905, Brazil}

\date{\today}

\begin{abstract}
This work investigates the use of dynamical decoupling to shield quantum discord from errors introduced by the environment. Specifically, a two-qubits system interacting with independent baths of bosons is considered. The initial conditions of the system were chosen as pure and mixed states, while the dynamical decoupling has been achieved by means of continuous fields. The effects of the temperature on the shielding of quantum discord is also studied. It is shown that although the quantum discord for particular initial states may be perfectly preserved over some finite time window in the absence of any protective field, the effectiveness of the dynamical decoupling with continuous fields depends essentially on the timescale required to preserve quantum discord. It is also shown that for these particular initial states the time for which the shielding of the quantum discord becomes effective decreases as the temperature increases.
\end{abstract}


\maketitle

\section{Introduction}
One of the major obstacles in building an efficient quantum computer is the presence of errors introduced by the environment. For over twenty years, different methods have been developed aiming to protect the quantum information, such as quantum error-correction codes \cite{shor95}, decoherence-free subspaces or subsystems \cite{zanardi97} and dynamical decoupling \cite{viola98, viola99, viola99b,viola04}. Among these strategies, the dynamical decoupling has the advantage of not requiring the use of extra qubits to encode the logical qubits, actively protecting quantum information by means of a sequence of ultrafast pulses or high frequency fields.

Quantum entanglement has been considered one of the main concepts concerning the measurement of quantum correlations. However, several alternative measures of quantum correlations have drawn considerable attention in the last few years. Among these plethora of new correlation measurements \cite{plet}, quantum discord \cite{ha, olivier} has been one of the most employed. Quantum discord (QD) is intimately connected to the entanglement of formation \cite{winter, law, LII}, the conditional entropy \cite{LII} and the mutual information. Moreover, quantum discord is related to many important protocols as the distribution of entanglement \cite{cavalcanti}, quantum locking \cite{lock}, entanglement irreversibility \cite{irr}, and many others \cite{qdother}.

For some specific initial conditions \cite{mazzola,maziero,pp1}, QD has the property of not being affected by the environment during an initial finite time. Such a finite time is known as \textit{transition time} because it delimits a sudden transition from classical to quantum decoherence regime. Recently, Luo \textit{et al.} \cite{luo} investigated the possibility of extending the \textit{transition time} {by means of a dynamical decoupling (DD) technique that uses} bang-bang pulses to control the decoherence of two qubits coupled to independent spin baths. Their results demonstrated that the \textit{transition time} can indeed be extended {by this kind of} protection.
In the present work, we theoretically investigate the protection of the QD through the application of {a different type of DD} technique that uses continuous fields, which is a more realistic scheme than using instantaneous pulses \cite {cont}. The interest of the scientific community has been increasing for this kind of protection and very recently it has been experimentally implemented to protect the quantum information in nitrogen vacancy systems \cite{exp, cai}. It has been shown that using this kind of protection, the coherence time can be prolonged about twenty times \cite{cai}. Furthermore, it can be applied to a wide variety of physical systems including trapped atoms and ions, quantum dots and nitrogen-vacancy centers in diamond.

In order to probe the protection method, we solve the Redfield master equation taking into account the interaction of each qubit with their own boson bath. We study the effects caused by temperature and two kinds of initial conditions: pure and mixed states. We show that protecting QD by means of DD technique can be tricky depending on the initial condition of the system and on the timescale required to preserve QD. Contrarily to entanglement and fidelity, the protection of quantum discord by means of DD is not efficient for some initial states with special symmetries before the \textit{transition time}. However, if one is interested in the long-term behavior of QD, then the DD protective scheme should be considered and the control field should be turned on well before the \textit{transition time}. Such a result is very different from the one obtained through instantaneous pulses approach \cite{luo}. This difference occurs because instantaneous pulses applied to the system do not break the symmetry of the reduced density matrix, which is an essential ingredient to keep QD intact before the \textit{transition time} \cite{mazzola,maziero,pp1}. On the other hand, a continuous field applied to the system induces a dynamics that breaks the symmetry of the reduced density matrix, thereby inhibiting the sudden transition.
\section{Dynamical decoupling}
The Hamiltonian for a two-qubit system and the environment can be written as

\begin{eqnarray}
H=H_{0}+H_{E}+H_{\rm int},\label{H}
\end{eqnarray}
where $H_{0}$ is the environment-free Hamiltonian, which includes the qubits Hamiltonian and the time-dependent fields applied to protect the quantum discord. $H_{E}$ is the Hamiltonian describing the environment and the Hamiltonian $H_{\rm int}$ takes into account the interaction between the qubits and the environment. Here,
\begin{equation}
H_E=\sum_{i=1}^2\sum_{k}\omega^{i}_k{a^{i}_k}^\dagger a^{i}_k\label{he}
\end{equation}
represents two independent baths of harmonic oscillators (one for each qubit) with $\omega^{i}_k$ given by the frequency of the $k$th normal mode of the thermal bath and $a^{i}_k$ $({a^{i}_k}^\dagger)$ is the usual annihilator (creator) operator for the $i$th qubit, \textit{i.e.}, $a^i_k$ (${a^{i}}_k^\dagger$) is the operator that annihilates (creates) a bath quantum in the $k$th mode of the $i$th qubit. We used units of $\hbar=1$ in Eq.~(\ref{he}). Henceforth, we adopt this convention.

The interaction Hamiltonian $H_{\rm int}$ describing two qubits coupled to their baths is written as
\begin{eqnarray}
H_{\rm int}={\cal L}_{1}{\sigma}^{(1)}_z+ {\cal L}_{2}{\sigma}^{(2)}_z,\label{Hint}
\end{eqnarray}
where ${\cal L}_{i}$, for $i=1,2$, are operators that act on the environmental Hilbert space. $\sigma_{z}^{(i)}$ are the Pauli $z$ matrix acting on $i$th qubit . This model is called a dephasing noise and it is a typical class of errors in quantum dots \cite{leo,tow} and Josephson junction \cite{tow}.  Explicitly, ${\cal L}_i=B_{i}+B_i^\dagger$, $B_{i}=\sum_k g_k^{i}a_k^{i}$, where $g_k^i$ is a complex coupling constant for $k$th normal mode with frequency dimension.

For the sake of simplicity, we assume there exist a global, static control field that exactly cancels the intrinsic system Hamiltonian, \textit{i.e.}, $H_0$ represents only the active protection and it is given by a continuous applied field defined as
\begin{eqnarray}
H_0=n_{x}\omega\left( {\sigma}^{(1)}_x+{\sigma}^{(2)}_x\right),\label{h0}
\end{eqnarray}
where $\omega =2\pi/t_{c}$, $t_c$ is the period of $U_0(t)=\exp(-iH_0t)$, and $n_x$ is an integer number.
The general prescription of the dynamical decoupling \cite{viola98,facchi05} is based on finding a controlled unitary operation such that
\begin{eqnarray}
\int ^{t_{c}} _{0} U^{\dagger}_{0}(t)H_{\rm int}U_{0}(t) dt=0.\label{integral}
\end{eqnarray}
This is a sufficient condition to shield the effect of the environment over the system and, as exposed in Ref. \cite{first}, perfect shielding can be reached when $n_x$ tends to infinity.

\section{Master equation}
In this work, we assume that the interaction between the qubit and its
environment is sufficiently weak such that linear-response theory holds. In the interaction picture, the master equation is written as
\begin{eqnarray}
\frac{d\rho _{I}(t)}{dt}=-\int^{t}_{0}dt^{\prime} {\rm Tr}_{E}\left\{{\left[H_{I}(t),\left[H_{I}(t^{\prime}),\rho _{E}(0)\rho _{I}(t)\right]\right]}\right\},\label{master}
\end{eqnarray}
while $H_{I}(t)$ reads
\begin{eqnarray}
H_{I}(t)=U_0^{\dagger}(t)U^{\dagger}_{E}(t)H_{\rm int}U_{E}(t)U_0(t),
\end{eqnarray}
where
\begin{eqnarray}
U_{E}(t)=\exp\left(-{\rm i}\frac{H_{E}t}{\hbar}\right),
\end{eqnarray}
and
\begin{eqnarray}
U_{0}(t)=\exp\left(-{\rm i}\frac{H_{0}t}{\hbar}\right).
\end{eqnarray}
In Eq.~(\ref{master}), $\rho _{E}(0)$ is the initial density matrix of the thermal bath, which is chosen as the Feynman-Vernon state,
\begin{eqnarray}
\rho _{E}(0)=\frac{1}{Z}\exp(-\beta H_{E}),\label{rhoE}
\end{eqnarray}
where $Z$ is the partition function given by
\begin{eqnarray}
Z={\rm Tr}_{E}\left[\exp(-\beta H_{E})\right],\label{Z}
\end{eqnarray}
and $\beta=1/(k_BT)$, where $k_B$ is the Boltzmann constant and $T$ is the temperature of the environment. Defining $U_E^\dagger(t) {\cal L}_i U_E(t)\equiv \tilde{\cal L}_i(t)$ and $\Lambda_i(t)\equiv U_{0}^\dagger(t) \sigma_z^{(i)} U_0(t)$, we can write the $H_I$ Hamiltonian in the interaction picture as follows,
\begin{eqnarray}
H_I(t)=\Lambda_1(t) \tilde{\cal L}_1(t) + \Lambda_2(t) \tilde{\cal L}_2(t). \label{hi}
\end{eqnarray}
Thus, substituting Eq.~(\ref{rhoE}) and Eq.~(\ref{hi}) into the master equation Eq.~(\ref{master}), we obtain
\begin{eqnarray}
\frac{d\rho_I(t)}{dt}&=&\sum_{i=1}^2\int_0^tdt^\prime {\cal D}_i(t,t^\prime)[\Lambda_i(t),\rho_I(t)\Lambda_i(t^\prime)]\nonumber\\
&+&\int_0^tdt^\prime {\cal D}_i^\ast(t,t^\prime)[\Lambda_i(t^\prime)\rho_I(t),\Lambda_i(t)],
\end{eqnarray}
where ${\cal D}_i(t,t^\prime)={\cal I}_i(t-t^\prime) + {\cal T}_i(t-t^\prime)$ with
\begin{eqnarray}
{\cal I}_i(t-t^\prime)&=&{\rm {Tr}}_E\left\{{\tilde B}_i(t){\tilde \rho}_E(t) {\tilde B}_i^\dagger(t^\prime)\right\}\nonumber\\
&=&\sum_k|g_k^i|^2n_k^i\exp[-{\rm i}\omega_k^i(t-t^\prime)]
\end{eqnarray}
and
\begin{eqnarray}
{\cal T}_i(t-t^\prime)&=&{\rm {Tr}}_E\left\{{\tilde B}^\dagger_i(t){\tilde \rho}_E(t) {\tilde B}_i(t^\prime)\right\}\nonumber\\
&=&\sum_k|g_k^i|^2(n_k^i+1)\exp[-{\rm i}\omega_k^i(t-t^\prime)],
\end{eqnarray}
with ${\tilde B}_i(t)=U_E^\dagger(t) B_i U_E(t)$, ${\tilde \rho}_E(t)=U_E^\dagger(t) \rho_E(0) U_E(t)$, and $n_k^i$ is the average occupation number of $k$th mode of the $i$th qubit, $n_k^i=1/[\exp(\beta\omega_k^i -1)]$. Since each qubit is subjected to identical environments, the $i$ index of $g_k^i$, $n_k^i$ and $\omega_k^i$ can be omitted. Thus, defining the spectral function as $J(\omega)\equiv\sum_k|g_k|^2\delta(\omega-\omega_k)$, we can replace the above summations by the following integrals
\begin{equation}
{\cal I}(t)=\int_0^\infty d\omega J(\omega)n(\omega)\exp(-{\rm i}\omega t),
\end{equation}
and
\begin{equation}
{\cal T}(t)=\int_0^\infty d\omega J(\omega)\exp({\rm i}\omega t)[n(\omega)+1].
\end{equation}
We assume that the environment has an ohmic spectral density, $J(\omega)=\eta\omega\exp(-\omega/\omega_c)$, where $\omega_c$ is a cut-off frequency and $\eta$ is a damping constant. Therefore, we can explicitly evaluate the above integrals which yield
\begin{eqnarray}
{\cal D}(t,t^\prime)&=&\frac{\eta\omega_c^2}{[1+i\omega_c(t-t^\prime)]^2}\nonumber\\
&&\!\!\!\!\!\!\!\!\!\!\!\!\!\!\!\!\!\!\!+\frac{2\eta}{\beta^2}{\rm{Re}}\left\{\Psi^{(1{\rm{st}})}(1+1/(\beta\omega_c)-i(t-t^\prime)/\beta) \right\}\!\!,\label{Dttl}
\end{eqnarray}
where $\Psi^{(1{\rm{st}})}$ is the first polygamma function. The first term of the righthand side of Eq.~(\ref{Dttl}) represents the vacuum, while the second term accounts for the effects of finite temperature. In particular, for $T=0$ we have ${\cal D}(t,t^\prime)={\eta\omega_c^2}/{[1+i\omega_c(t-t^\prime)]^2}$.
\section{Fidelity, Concurrence, and Quantum Discord}
\commentold{To elucidate the peculiar aspects involving the protection of the quantum discord, we compare the later dissipative dynamics with entanglement of formation and fidelity.}
In this section, we present the definition of three distinct measures for quantum systems.
\subsection{Superfidelity}
To calculate the dissipative dynamics fidelity, we use an alternative measure known in the literature as \textit{superfidelity} \cite{fidelity}. We calculate the superfidelity between two distinct (possibly mixed) density matrix state, $\rho(t)$ and $\sigma(t)$.  Here, $\rho(t)$ represents the open quantum system dynamics, where the two qubits interact with the environment and $\sigma(t)$ gives the ideal dynamics, whose interaction with the environment is disabled.
The superfidelity is a function of linear entropy and the Hilbert-Schmidt inner product between the given states. Remarkably, the superfidelity is jointly concave and satisfies all Jozsa's axioms. Explicitly it is given by:
\begin{equation}
F(\rho,\sigma)={\rm{Tr}}(\rho\sigma)+\sqrt{1-{\rm{Tr}}(\rho^2)}\sqrt{1-{\rm{Tr}}(\sigma^2)},
\end{equation}
and it is valid for pure or mixed states.
\subsection{Concurrence}
To measure the entanglement between the two qubits, we use the well known concurrence \cite{conc}. It is defined as the maximum between zero and $\Lambda(t)$,
\begin{equation}
\Lambda(t)=\lambda_1-\lambda_2-\lambda_3-\lambda_4,
\end{equation}
where $\lambda_1\ge\lambda_2\ge\lambda_3\ge\lambda_4$ are the square roots of the eigenvalues of the matrix $\rho(t)\sigma_y\otimes\sigma_y\rho^\ast(t)\sigma_y\otimes\sigma_y$, where $\rho^\ast(t)$ is the complex conjugation of $\rho(t)$ and $\sigma_y$ is the $y$ Pauli matrix.

\subsection{Quantum Discord}
The quantum discord was independently defined by Handerson and Vedral \cite{ha} and Ollivier and Zurek \cite{olivier} about ten years ago. It was proposed as a measure of the quantum correlations and, given two subsystem $A$ and $B$, it is defined as
\begin{equation}
\delta_{AB}^\leftarrow=I_{AB}-J_{AB}^\leftarrow
\end{equation}
where $I_{AB}=S_A + S_B - S_{AB}$ is the mutual information, with $S_A=-{\rm{Tr}}(\rho_A\log\rho_A)$ given by the von-Neumann entropy of the subsystem $A$ and similarly to $B$ and $AB$. $J_{AB}^\leftarrow$ is the classical correlation, which can be explicitly written as
\begin{equation}
J_{AB}^\leftarrow=\max_{\{\Pi_k\}}\left[ S_A - \sum_k p_k S(\rho_{A|k})\right],
\end{equation}
where $\rho_{A|k} = {\rm{Tr}}_B\left(\Pi_k \rho_{AB} \Pi_k\right)/{\rm {Tr}}_{AB}\left(\Pi_k \rho_{AB} \Pi_k\right)$ is the reduced state of $A$ after obtaining the outcome $k$ in $B$.
Here, $\{\Pi_k\}$ is a complete set of positive operator valued measures that results in the outcome $k$ with probability $p_k$. The quantum discord measures the amount of the mutual information that is not locally accessible \cite{demon, LII} and generally is not symmetric, \textit{i.e.}, $\delta_{AB}^\leftarrow\ne\delta_{BA}^\leftarrow$.

\begin{figure}[htbp]
\begin{center}
\includegraphics[width=.45\textwidth]{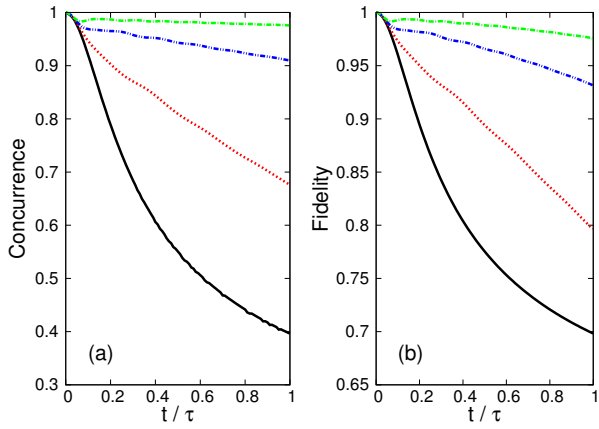} {}
\end{center}
\caption{(Color Online) Concurrence and fidelity as a function of time considering a pure initial state $\rho^0_{AB}(0)$ and $T=0$K. (a) The solid (black) line gives the concurrence when the control field is turned off and the dotted (red), double dot-dashed (blue), and dot-dashed (green) lines give the concurrence when the control fields are turned on at $t=0$, using $n_x = 2, 3, 4$ respectively. (b) The solid (black) line gives the fidelity when the control field is turned off and the dotted (red), double dot-dashed (blue), and dot-dashed (green) lines give the fidelity when the control fields are turned on at $t=0$, using $n_x = 2, 3, 4$ respectively.}\label{fig1}
\end{figure}

\section{Dissipative Dynamics}
In order to illustrate the singular features involving the protection of the quantum discord, we study two kinds of initial conditions, pure states and the mixed ones. Additionally, we consider the effects of finite temperatures on the quantum discord. The quantum discord dynamics is compared to the concurrence and the fidelity of the quantum systems. In the following calculations, we have set the parameters of  Eq.~(\ref{Dttl}) to $\eta=1/16$ and $\omega_c=2\pi$.
\begin{figure}[htbp]
\begin{center}
\includegraphics[width=.45\textwidth]{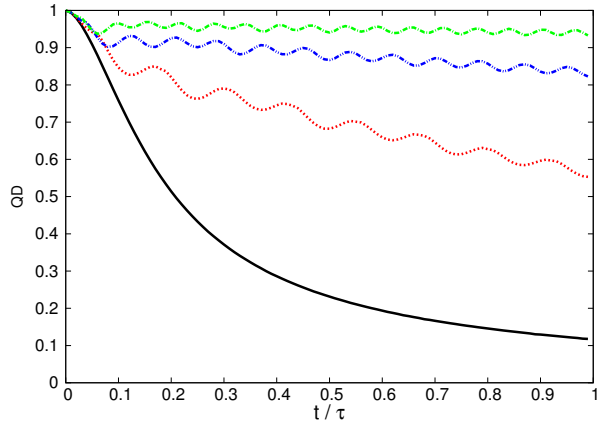} {}
\end{center}
\caption{(Color Online) Quantum Discord as a function of time considering a pure initial state $\rho^0_{AB}(0)$ and $T=0$K. The solid (black) line gives the quantum discord when the control field is turned off and the dotted (red), double dot-dashed (blue), and dot-dashed (green) lines give the quantum discord when the control fields are turned on at $t=0$, using $n_x = 2, 3, 4$ respectively.}\label{fig2}
\end{figure}
\subsection{Pure-state initial condition}
In order to analyze the dissipative dynamics for pure states, we choose for the initial condition a maximal quantum correlated state {$\rho^0_{AB}(0)=|\Phi^+\rangle\langle\Phi^+|$ with $|\Phi^+\rangle=\frac{1}{\sqrt{2}}(|00\rangle+|11\rangle)$}.

Figure (\ref{fig1}) shows the concurrence (a) and the fidelity (b) dynamics when the control fields are turned off ($n_x=0$) and on ($n_x=2, 3, 4$) for a zero temperature environment, $T=0$K. The solid line corresponds to $n_x=0$ in Eq.~(\ref{h0}) and the dashed, dotted, and dot-dashed lines give the concurrence and the fidelity when the control fields are turned on at $t=0$ with $n_x = 2, 3, 4$, respectively. When the protective field amplitude increases both the fidelity and the concurrence increase \cite{note}.
\begin{figure}[htbp]
\begin{center}
\includegraphics[width=.45\textwidth]{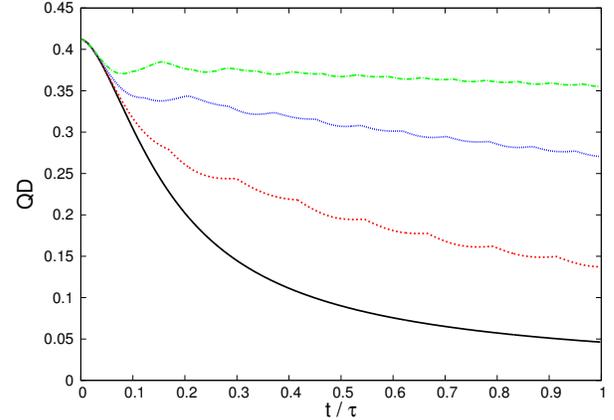}
\end{center}
\caption{(Color Online) Quantum Discord as a function of time considering a mixed initial state $\rho^1_{AB}(0)$ and $T=0$K. The solid (black) line gives the quantum discord when the control field is turned off and the dotted (red), double dot-dashed (blue), and dot-dashed (green) lines give the quantum discord when the control fields are turned on at $t=0$, using $n_x = 2, 3, 4$ respectively.}\label{mixed}
\end{figure}

In Figure (\ref{fig2}), quantum discord is plotted considering the same parameters used for fidelity and concurrence. The solid line corresponds to $n_x=0$ in Eq.~(\ref{h0}) and the dashed, dotted, and dot-dashed lines give the quantum discord when the control fields are turned on at $t=0$ with $n_x = 2, 3, 4$, respectively. For the initial pure state case, the concurrence, fidelity, and quantum discord presents a similar behavior, \textit{i.e.}, as the amplitude of the protective field increases, the protection of the quantum information is enhanced \cite{note}. It is important to emphasize that, despite the restrictive initial state, the protection of fidelity, concurrence, and quantum discord is reached for all pure initial states qualitatively in the same way, that is, the higher the amplitude of the protective field, the better the protection. As shown in the following, the quantum discord does not necessarily behaves in the same qualitative way as fidelity and concurrence with respect to the application of the protective field for mixed states.

\subsection{Mixed-state initial condition}

Consider as initial condition a mixed state given by $\rho^1_{AB}(0)=0.5(|\psi_1\rangle\langle\psi_1|+|\psi_2\rangle\langle\psi_2|)$, where $|\psi_1\rangle=\frac{1}{\sqrt{4}}(|00\rangle+|11\rangle+|01\rangle+|10\rangle)$ and $|\psi_2\rangle=\frac{1}{\sqrt{2}}(|00\rangle-|11\rangle)$.
Fig.~(\ref{mixed}) shows the dynamics of QD as function of time, considering $n_x = 0, 2, 3, 4$.
One can notice that similarly to the pure initial state case, the control field is efficient in protecting QD. In brief, the protection protocol has the same efficiency for the initial state $\rho^1_{AB}(0)$ as the one found for pure initial states. Such a characteristic is also observed for the great majority of initial mixed-states.

\begin{figure}[htbp]
\begin{center}
\includegraphics[width=.45\textwidth]{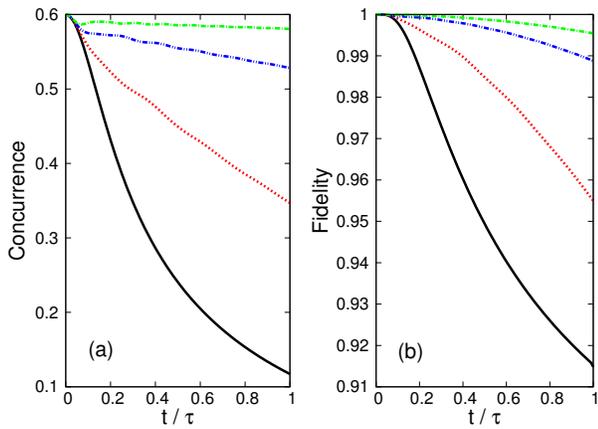} {}
\end{center}
\caption{(Color Online) Concurrence and Fidelity as a function of time considering a mixed initial state $\rho^2_{AB}(0)$ and $T=0$K. (a) The solid (black) line gives the concurrence when the control field is turned off and the dotted (red), double dot-dashed (blue), and dot-dashed (green) lines give the concurrence when the control fields are turned on at $t=0$, using $n_x = 2, 3, 4$ respectively. (b) The solid (black) line gives the fidelity when the control field is turned off and the dotted (red), double dot-dashed (blue), and dot-dashed (green) lines give the fidelity when the control fields are turned on at $t=0$, using $n_x = 2, 3, 4$ respectively.}\label{fig3}
\end{figure}

On the other hand, the scenario changes if the initial condition is given by some special states, for instance, $\rho^2_{AB}(0)=0.8|\Phi\rangle\langle\Phi|+0.2|\Psi\rangle\langle\Psi|$, where $|\Phi\rangle=\frac{1}{\sqrt{2}}(|00\rangle+|11\rangle)$ and $|\Psi\rangle=\frac{1}{\sqrt{2}}(|01\rangle+|10\rangle)$.
In Fig. (\ref{fig3}), we show the concurrence and the fidelity dynamics for some frequencies of the protective field. There is no qualitative difference involving the protection of the concurrence and the fidelity: the greater the amplitude of the protective field, the better the protection. Nevertheless, we observe a different behavior when dealing with the quantum discord.

\begin{figure}[htbp]
\begin{center}
\includegraphics[width=.45\textwidth]{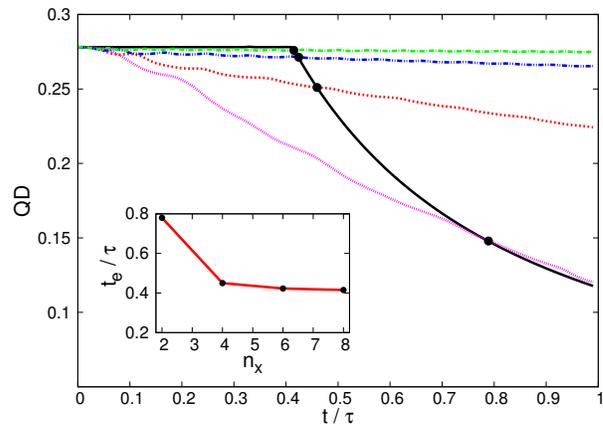} {}
\end{center}
\caption{(Color Online) Quantum Discord as a function of time considering a mixed initial state $\rho^2_{AB}(0)$ and $T=0$K. The solid (black) line gives the quantum discord when the control field is turned off and the closed points (magenta), dotted (red), double dot-dashed (blue), and dot-dashed (green) lines give the quantum discord when the control fields are turned on at $t=0$, using $n_x = 1, 2, 3, 4$ respectively. In the inset we plot the time for efficiency as a function of $n_x$.}\label{fig4}
\end{figure}

In Fig.~(\ref{fig4}), it can be noticed an abrupt decrease, the sudden transition, in the quantum discord around $t=0.4\tau$ for the non-protected case. Up to this sudden transition, the results obtained with the protection turned on at $t=0$ are less efficient in maintaining the quantum discord than with the protection turned off. Actually, for $t\le0.4\tau$, the protective dynamics is equivalent to the non-protected one only in the limit that $n_x$ goes to infinity. However, there is a finite time when the protected quantum discord surpass that of the non-protected case, given by the crossing of the dotted lines with the solid line in Fig. (\ref{fig4}). The effectiveness of the protective field only occurs above these times $t_e(n_x)$, which are greater than the time of sudden transition of the quantum discord $t_e(n_x)\ge0.4\tau$. Thus, for this particular initial condition, the timescale needed to protect QD should be determinant to decide for the use of the DD protective scheme.
This peculiar aspect involving the protection of the quantum discord occurs because the plateau given in the free evolution exist for some special symmetries of the initial state \cite{maziero,mazzola}. The dynamical decoupling strategy is an active method of protection and needs to continuously modify the density matrix during the dynamics. On the other hand, the density matrix reaches its original and special symmetry just at some periodic times. These unavoidable density matrix modifications involving the protection by means of DD decreases the quantum discord, which reaches a lower value than the initial one. However, when the quantum discord dynamics reaches the sudden transition time, it starts to decrease, while the protective scheme becomes efficient. {It is important to emphasize the difference between our results and the ones obtained by the DD technique that employs instantaneous pulses \cite{luo}. If each pulse is considered to be applied instantaneously, the changes imposed to the system by this kind of pulses never alter the X structure of the initial reduced density matrix. Therefore, the sharp transition in the quantum discord is preserved, whereas the continuous decoupling technique presented here leads to a more gradual decay.}

\begin{figure}[htbp]
\begin{center}
\includegraphics[width=.45\textwidth]{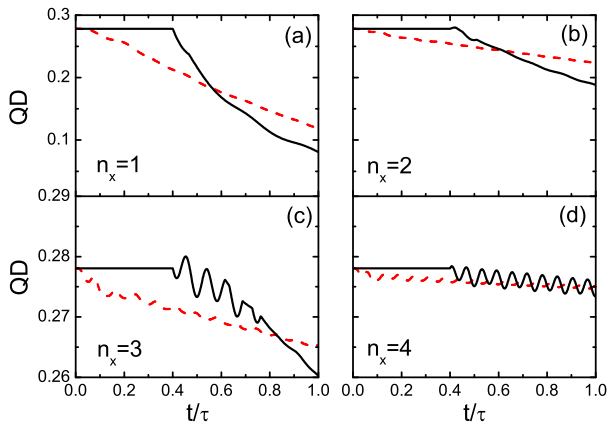} {}
\end{center}
\caption{(Color Online) Quantum Discord as a function of time considering a mixed initial state $\rho^2_{AB}(0)$ and $T=0$K. The solid (black) lines give the quantum discord when the control field is turned on at $t=0.4\tau$, while the dashed (red) curves correspond to the QD when the control fields is turned on at $t=0$. We probe both situations  by using $n_x = 1$ in (a), $n_x = 2$ in (b), $n_x = 3$ in (c), and $n_x = 4$ in (d).}\label{fignew}
\end{figure}

Based on the above results, one could be led to conclude that if the initial state has a special symmetry such that the QD presents a plateau for the free evolution of the system, then the best strategy would be to turn on the protective field only after the sudden transition time.
 However, as we show in the following, this may not be the best strategy. Figure~(\ref{fignew}) presents the evolution of the QD for the protective field turned on at different moments. The dashed curves probe the situation with the protective field turned on at $t=0$, while the solid curves probe the dynamics for a control field turned on at $t=0.4\tau$. It is clearly seen that if one intends to enhance the protection of QD in the long-term, one should turn on the protective field at $t=0$ and not wait until the sudden transition.

It is interesting to analyze the case of protection for completely unknown states, where
an initial state is chosen randomly. In this case, the density matrix will most likely never reach the necessary structure to result in a frozen discord \cite{pp1}. Thus, considering the long term evolution, the best strategy is to turn on the protection at $t=0$. Nevertheless, it is important to note that a state that presents a plateau in the QD is a mixture of Bell states. These states are a very important class of states and are used in many problems in quantum information theory \cite{pp2}. Furthermore,
experimental realization of these states has been observed in the context of quantum optics experiments
\cite{pp3}, nuclear magnetic resonance \cite{pp4} as well as in solid state physics \cite{pp5}.

\begin{figure}[htbp]
\begin{center}
\includegraphics[width=.45\textwidth]{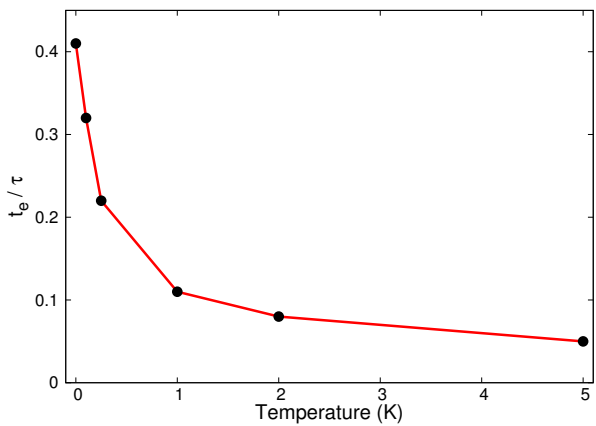} {}
\end{center}
\caption{(Color Online) The time for efficiency as a function of the temperature for $n_x=4$.}\label{figTemp}
\end{figure}

We now consider how a finite temperature environment influences the scenario of QD protection. The first relevant observation involving finite temperature environments is that the interaction with the environment becomes more deleterious. In Fig. (\ref{figTemp}), we study the temperature effects on the protection method. We take $n_x=4$ and present the time for effectiveness of the protective field as a function of the temperature. As one can observe, ${t_e}(nx)$ decreases when the temperature is increased. Therefore, for higher temperatures, the protective scheme becomes efficient sooner. This characteristic occurs because when the temperature increases, the density matrix coherence decays faster and, consequently, the time of sudden transition is reduced. Thus, the plateau of quantum discord is smaller and the effectiveness of the protective scheme is achieved sooner. We note that for $T=5$K the time for efficiency approaches $0.05\tau$.

\section{Summary}
In summary, we have investigated the protection of QD from errors introduced by the environment  by means of DD with continuous fields. We have compared two kinds of initial condition, pure and mixed states and also included finite temperature effects. It has been shown that, in contrast with fidelity and concurrence, the effectiveness of the protective scheme depends on the initial state and on the timescale required to protect QD. For initial states with a particular symmetry such that the QD presents a plateau in the free evolution of the system, one should not turn on the protective field if the timescale of interest is below the \textit{transition time}. However, if one is interested in the long term evolution of QD, the protective field should be turned on well before the  \textit{transition time}. Finally, we have verified that for these particular initial states, the greater the temperature, the faster the efficiency of the protective method is achieved, which evidences the relevance of the temperature in the protection of QD.

\begin{acknowledgments}
This work is supported by FAPESP and CNPq through the
National Institute for Science and Technology of Quantum
Information (INCT-IQ).

\end{acknowledgments}


\begin{thebibliography}{99}
\bibitem{shor95} P. W. Shor, Phys. Rev. A {\bf 52}, R2493 (1995); A. M. Steane, Phys. Rev. Lett. {\bf 77}, 793 (1996); A. R. Calderbank and P. W. Shor, Phys. Rev. A {\bf 54}, 1098 (1996); A. M. Steane, Proc. R. Soc. London A {\bf 452}, 2551 (1996); M. S. Byrd, L.-A. Wu, and D. A. Lidar, J. Mod. Opt. {\bf 10}, 2449 (2004), and references cited therein.
\bibitem{zanardi97} P. Zanardi and M. Rasetti, Phys. Rev. Lett. {\bf 79}, 3306 (1997); D. A. Lidar, I. L. Chuang, and K. B. Whaley, Phys. Rev. Lett. {\bf 81}, 2594 (1998); E. Knill, R. Laflamme, and L. Viola, Phys. Rev. Lett. {\bf 84}, 2525 (2000); P. E. M. F. Mendon\c{c}a, M. A. Marchiolli, and R. d. J. Napolitano, J. Phys. A: Math. Gen. {\bf 38}, L95 (2005), and references cited therein.
\bibitem{viola98} L. Viola and S. Lloyd, Phys. Rev. A {\bf 58}, 2733 (1998).
\bibitem{viola99} L. Viola, E. Knill, and S. Lloyd, Phys. Rev. Lett. {\bf 82}, 2417 (1999).
\bibitem{viola99b} L. Viola, S. Lloyd, and E. Knill, Phys. Rev. Lett. {\bf 83}, 4888 (1999).
\bibitem{viola04} L. Viola, J. Mod. Opt. {\bf 10}, 2357 (2004), and references cited therein.
\bibitem{facchi05} P. Facchi, S. Tasaki, S. Pascazio, H. Nakazato, A. Tokuse, and D. A. Lidar, Phys. Rev. A {\bf 71}, 022302 (2005).
\bibitem{plet} M. Horodecki, K. Horodecki, P. Horodecki, R. Horodecki, J. Oppenheim, A. Sen De, and U. Sen , Phys. Rev. Lett. {\bf 90}, 100402 (2003); M. Horodecki, P. Horodecki, R. Horodecki, J. Oppenheim, A. Sen, U. Sen, and B. Synak-Radtke, Phys. Rev. A {\bf 71}, 062307 (2005);
S. Luo, Phys. Rev. A {\bf 77}, 022301 (2008); J. Maziero, L. C. Celeri, and R. M. Serra, arXiv:1004.2082; K. Modi, T. Paterek, W. Son, V. Vedral, and M. Williamson, Phys. Rev. Lett. {\bf 104}, 080501 (2010).
\bibitem{ha} L. Henderson and V. Vedral, J. Phys. A {\bf 34}, 6899 (2001);
\bibitem{olivier} H. Ollivier and W. H. Zurek, Phys. Rev. Lett. {\bf 88}, 017901 (2001).
\bibitem{winter} M. Koashi and A. Winter, Phys. Rev. A {\bf 69}, 022309 (2004).
\bibitem{law} F. F. Fanchini, M. F. Cornelio, M. C. de Oliveira, A. O. Caldeira, Phys. Rev. A \textbf{84}, 012313 (2011).
\bibitem{LII} F. F. Fanchini, L. K. Castelano, M. F. Cornelio, M. C. de Oliveira, New J. Phys. \textbf{14}, 013027 (2012).
\bibitem{cavalcanti} V. Madhok and A. Datta, Phys. Rev. A \textbf{83}, 032323 (2011); D. Cavalcanti, L. Aolita, S. Boixo, K. Modi, M. Piani, and A. Winter, Phys. Rev. A \textbf{83}, 032324 (2011);
A. Streltsov, H. Kampermann, and D. Bruss, Phys. Rev. Lett. \textbf{108}, 250501 (2012); T. K. Chuan, J. Maillard, K. Modi, T. Paterek, M. Paternostro, and M. Piani, Phys. Rev. Lett. \textbf{109}, 070501 (2012).
\bibitem{lock} S. Boixo, L.  Aolita, D. Cavalcanti, K. Modi, A. Winter,  Int. J. Quant. Inf. {\bf 9}, 1643 (2011).
\bibitem{irr} M. F. Corn\'elio, M. C. de Oliveira, and F. F. Fanchini,  Phys. Rev. Lett. \textbf{107}, 020502 (2011).
\bibitem{qdother} A. Streltsov, H. Kampermann, and D. Bruss,  Phys. Rev. Lett. {\bf 106}, 160401 (2011); T. Werlang, C. Trippe, G. A. P. Ribeiro, and G. Rigolin,  Phys. Rev. Lett. {\bf 105}, 095702 (2010).
\bibitem{mazzola} L. Mazzola, J. Piilo, and S. Maniscalco,  Phys. Rev. Lett. {\bf 104}, 200401 (2010).
\bibitem{maziero} J. Maziero, L. C. Celeri, R. M. Serra, and V. Vedral, Phys. Rev. A {\bf 80}, 044102 (2009).
\bibitem{pp1} B. You, L.-X. Cen, Phys. Rev. A \textbf{86}, 012102 (2012).
\bibitem{luo} Da-Wei Luo, Hai-Qing Lin, Jing-Bo Xu, and Dao-Xin Yao, Phys. Rev. A {\bf 84}, 062112 (2011).
\bibitem{cont}  F. F. Fanchini, J. E. M. Hornos, and R. d. J. Napolitano, Phys. Rev. A {\bf 75}, 022329 (2007); F. F. Fanchini and R. d. J. Napolitano, Phys. Rev. A {\bf 76}, 062306 (2007).
\bibitem{exp} X. Xu, et al., arXiv: 1205.1307.
\bibitem{cai} J.-M. Cai, et al., arXiv: 1111.0930.
\bibitem{leo} F. F. Fanchini, L. K. Castelano, and A. O. Caldeira, New J. of Phys. {\bf 12}, 073009 (2010).
\bibitem{tow} M. Thorwart and P. Hanggi, Phys. Rev. A {\bf 65} 012309 (2001).
\bibitem{first} F. F. Fanchini, J. E. M. Hornos, and R. d. J. Napolitano, Phys. Rev. A {\bf 75}, 022329 (2007).
\bibitem{fidelity} P. E. M. F. Mendonca, R. d. J. Napolitano, M. A. Marchiolli, C. J. Foster, and Y. C. Liang, Phys. Rev A. {\bf 78}, 052330 (2008); J. A. Miszczak, Z. Pucha la, P. Horodecki, A. Uhlmann, and K. Zyczkowski, Quantum Inf. Comput. {\bf 9}, 0103 (2009).
\bibitem{conc} W. K. Wootters, Phys. Rev. Lett. {\bf 80}, 2245 (1998).
\bibitem{demon} W. H. Zurek, Phys. Rev. A \textbf{67}, 012320 (2003).
\bibitem{note} We are not considering the anti-Zeno effect, where the protective field could decrease both the fidelity and the concurrence.

\bibitem{pp2} M. Zwerger, W. Duer, H. J. Briegel, Phys. Rev. A  \textbf{85}, 062326
(2012); A. Acin, N. Brunner, N. Gisin, S. Massar, S. Pironio, and V. Scarani, Phys.
Rev. Lett. \textbf{98}, 230501 (2007); C.H. Bennett, G. Brassard, C. Crepeau, R.
Jozsa, A. Peres, and W. Wootters, Phys. Rev. Lett. \textbf{70}, 1895 (1993).

\bibitem{pp3} J.-S. Xu. \textit{et al.}, Nat. Commun. 1, 7 (2010); M. F. Cornelio
\textit{et al.}, arXiv:1203.5068.

\bibitem{pp4} R. Auccaise \textit{et al.}, Phys. Rev. Lett. \textbf{107}, 140403
(2011);

\bibitem{pp5} X. Rong \textit{et al.}, arXiv:1203.3960.


\end{thebibliography}
\end{document}